\documentclass[a4paper,fleqn,usenatbib]{mnras}

\usepackage[T1]{fontenc}
\usepackage{ae,aecompl}
\usepackage{graphicx}
\usepackage{amsmath}
\usepackage{amssymb}
\usepackage{multirow}

\title[MRI in magnetically polarized discs]{Magneto-rotational instability in magnetically polarized discs}

\author[Oscar M. Pimentel et al.]{Oscar M. Pimentel$^{1,2}$\thanks{oscar.pimentel@correo.uis.edu.co}, P. Chris Fragile$^{2,3}$, 
F. D. Lora-Clavijo$^{1}$, Bridget Ierace$^2$,
\newauthor
and Deepika Bollimpalli$^{2,4}$
\\
$^{1}$Escuela de F\'isica, Universidad Industrial de Santander, A. A. 678, Bucaramanga 680002, Colombia\\
$^{2}$Department of Physics and Astronomy, College of Charleston, Charleston, SC 29424, USA\\
$^{3}$Kavli Institute for Theoretical Physics, University of California Santa Barbara, Santa Barbara, CA 93106, USA\\
$^{4}$Nicolaus Copernicus Astronomical Center, ul. Bartycka 18, PL 00-716 Warsaw, Poland
}

\date{Accepted XXX. Received YYY; in original form ZZZ}

\pubyear{2021}

\begin{document}
\label{firstpage}
\pagerange{\pageref{firstpage}--\pageref{lastpage}}
\maketitle

\begin{abstract}

The magneto-rotational instability (MRI) is the most likely mechanism for transportation of angular momentum and dissipation of energy within hot, ionized accretion discs. This instability is produced through the interactions of a differentially rotating plasma with an embedded magnetic field. Like all substances in nature, the plasma in an accretion disc has the potential to become magnetically polarized when it interacts with the magnetic field. In this paper we study the effect of this magnetic susceptibility, parameterized by $\chi_m$, on the MRI, specifically within the context of black hole accretion. We find from a linear analysis within the Newtonian limit that the minimum wavelength of the first unstable mode and the wavelength of the fastest growing mode are shorter in paramagnetic ($\chi_m>0$) than in diamagnetic ($\chi_m<0$) discs, all other parameters being equal. Furthermore, the magnetization parameter (ratio of gas to magnetic pressure) in the saturated state should be smaller when the magnetic susceptibility is positive than when it is negative. We confirm this latter prediction through a set of numerical simulations of magnetically polarized black hole accretion discs. We additionally find that the vertically integrated stress and mass accretion rate are somewhat larger when the disc is paramagnetic than when it is diamagnetic. If astrophysical discs are able to become magnetically polarized to any significant degree, then our results would be relevant to properly interpreting observations.

\end{abstract}


\section{Introduction}
\label{sec:intro}

Matter falling into a black hole provides one of the most efficient mechanisms to power the energetic emissions of quasars \citep{Wyithe03,Reeves05} and active galactic nuclei (AGN), which emit intense electromagnetic radiation across a broad range of frequencies \citep{Rees84,Reynolds14}. This mechanism likely also powers high-energy astrophysical sources like X-ray binaries \citep{Remillard06} and gamma ray bursts (GRBs) \citep{Piran99}. In addition, black hole accretion has been used as a laboratory to test General Relativity in the strong field regime \citep{Broderick14}. Observations of  the hot plasma moving in the vicinity of the black hole are now possible due to the Event Horizon Telescope (EHT), which recently imaged the supermassive black hole at the center of the galaxy M87 \citep{EHT19}.

The transfer of angular momentum, which is required to allow matter to reach the black hole, is most likely powered by the magneto rotational instability \citep[MRI;][]{Balbus91,Balbus98}. The MRI uses the differential rotation of the fluid, in conjunction with magnetic tension forces, to drive turbulence, which facilitates the transport of angular momentum and dissipation of energy. Magnetic fields in accretion discs likely also play important roles in powering jets \citep{Blandford77,McKinney06}, energizing coronae \citep{Haardt91,Jiang19}, and driving winds \citep{Blandford82}. In short, magnetic fields are a crucial component of black hole astrophysics.

An often overlooked aspect of accretion is that the plasma in the disc is itself composed of many tiny magnetic dipoles at the microscopic level, generated either by spinning and orbiting electrons bound within atoms or by free electrons orbiting around ambient magnetic field lines. Those dipoles have the potential to collectively align with (paramagnetism) or against (diamagnetism) the ambient magnetic field. The net result of this effect can be to enhance or diminish the effective field strength. 

Diamagnetism resulting from electrons gyrating around magnetic field lines has the effect of screening the external field. This effect is larger when the energy density of the injected particles is high compared to the energy density of the initial magnetic field \citep{ginzburg1969developments,singal1986magnetization}. One of the consequences of this kind of diamagnetism is that the system does not necessarily evolve to equipartition but rather to an equilibrium state that depends on the injection parameters \citep{bodo1992diamagnetic}. Within the context of accretion disc theory, \citet{Devlen07,Devlen20} show that this type of diamagnetism enhances the growth rate of the MRI and shifts the unstable region to shorter wavelengths. Otherwise, diamagnetism and paramagnetism have not generally been considered in accretion processes.

The main goal of this work is to explore the effect of the magnetic polarization produced by bound electrons in the generation and evolution of the MRI within an accretion disc. Here, the orbital motion of the electrons gives rise to the diamagnetism, so it should be a universal phenomenon. Although the diamagnetic susceptibility is usually weak, it could be important for high-density plasmas with large magnetic fields \citep{ray2001atomic}. These physical conditions are possibly found near the surface of a neutron star, where the magnetic field can be in the range of $10^{12-15}$ G and the mass density is $\sim 10^{13}$ g/cm$^{3}$. Therefore, a neutron star tidally disrupted by a black hole could be an astrophysical scenario where this diamagnetic behaviour becomes relevant in an accretion disc.

Paramagnetism, on the other hand, arises from the alignment of the spin magnetic moments of unpaired electrons with the direction of the ambient magnetic field. The degree of magnetic polarization in this case depends on the ratio of the energy of the magnetic dipole in the field to the thermal energy \citep{beaurepaire2001magnetism}. When this ratio is $\ll 1$ the degree of paramagnetic susceptibility follows the Curie law, thus decreasing with increasing temperature. Since the inner accretion disc in low-mass X-ray binaries reaches temperatures in the range of $10^{6-7}$ K, paramagnetic effects should be small according to the Curie law. Nevertheless, for the same range of temperatures and magnetic fields $\gtrsim10^{10}$G, the magnetization deviates from the Curie law and can even reach a saturation state for high enough magnetic fields. The relevance of paramagnetism becomes less clear in this scenario. Such conditions may be reached in accretion discs in black hole-neutron star binary systems, or any other scenario where the energy of the magnetic dipole in the magnetic field is comparable to the thermal energy of the plasma.

In order to study the behavior of a magnetically polarized fluid in general relativity, it is necessary to first introduce the effect of the magnetization vector into the energy momentum tensor. A suitable expression for the appropriately modified tensor has been obtained by many authors following two independent paths. The first is motivated by well-developed microscopic models \citep{groot1972foundations, Maugin:1978tu, felderhof2004relativistic}, while the other is obtained from the variational principle \citep{caldarelli2009dyonic, Chatterjee15, pu2016bjorken}. Both formulations are axiomatically equivalent, though the latter is much simpler. 

Using the resulting energy-momentum tensor and the conservative variables from the Valencia formulation \citep{banyuls1997numerical, 2006ApJ...637..296A}, we generalized the equations of relativistic magnetohydrodynamics to introduce the effects of magnetic susceptibility \citep{Pimentel20181}. Later, we presented the first family of magnetically polarized equilibrium tori around a Kerr black hole \citep{Pimentel20182}, which was obtained by following the Komissarov approach \citep{Komissarov06}. We also analyzed the effects of diamagnetism and paramagnetism on the evolution of the relativistic Kelvin-Helmholtz instability \citep{Pimentel2019}. In the linear regime we found that the growth rate of the instability and the magnetic field amplification are higher when the fluid is paramagnetic than when it has diamagnetic properties. In the non-linear regime we found that the effect of the magnetic susceptibility in the amplification of the magnetic energy is more relevant when the initial magnetic field is small. This is an interesting result for accretion processes since the saturation of the MRI may be related to parasitic instabilities of the Kelvin-Helmholtz and tearing-mode type \citep{goodman1994parasitic, pessah2009saturation}.

Motivated by the fact that all substances have the potential to become magnetically polarized, we feel it is fruitful to explore the effect of diamagnetism and paramagnetism on the growth of the MRI and the subsequent accretion process around black holes. In this paper, we redo the linear stability analysis of the MRI accounting for the effects of magnetic susceptibility. We compute the minimum wavelength of the unstable modes and the fastest growing mode in the vertical direction for diamagnetic and paramagnetic rotating fluids.  
We then confirm our findings through a set of global, general relativistic magnetohydrodynamic (GRMHD) simulations of accretion discs with differing levels of magnetic polarization.

\section{Linear analysis of the MRI in a magnetically polarized disc}
\label{sec:analysis}

The dynamics of a relativistic fluid in the MHD approximation are described by the following equations: energy-momentum conservation, $\nabla_{\mu}T^{\mu\nu}=0$; mass conservation, $\nabla_{\mu}(\rho u^{\mu})=0$; and magnetic induction, $\nabla_\mu(u^{\mu}b^{\nu}-b^{\mu}u^{\nu})=0$; where $u^{\mu}$ is the fluid four-velocity, $b^{\mu}$ is the magnetic field four-vector, and $\rho$ is the rest mass density in a co-moving reference frame. The energy-momentum tensor for a magnetically polarized fluid has the form \citep{Huang10, Chatterjee15}
\begin{align}
T^{\mu\nu}=&\left[\rho h+(1-\chi)b^2\right]u^{\mu}u^{\nu}+\left[p+\frac{1}{2}(1-2\chi)b^{2}\right]g^{\mu\nu} \nonumber \\
&-(1-\chi)b^{\mu}b^{\nu},
\label{energy-mometum_tensor}
\end{align}
where $h=1 + \epsilon + p/\rho$ is the enthalpy, $\epsilon$ is the specific internal energy, $p$ is the gas pressure, $g^{\mu\nu}$ is the metric tensor, and $\chi \equiv \chi_m/(1+\chi_m)$, $\chi_m$ being the magnetic susceptibility of the fluid. Negative values of $\chi_m$ describe diamagnetic materials, while positive values correspond to paramagnetic ones.

It is worth mentioning that typical values of $\chi_m$ are small ($\ll 1$); for example, the magnetization of electrons in ordinary atoms is rather weak. Nevertheless, in the crust of old radio pulsars where the electron's thermal energy is expected to be smaller than the spacing of the Landau levels, magnetic susceptibility can grow and undergo large de Haas-van Alphen oscillations \citep{blandford1982magnetic}. This effect can result in domain formation, with possible indirect observational consequences \citep{suh2010magnetic, wang2016diamagnetic}. However, the degree of magnetic polarization within accretion discs is still unknown, especially in strong magnetic field scenarios. Consequently, our present work lies in the theoretical realm, and the results are mainly focused on characterizing the general effects of susceptibility on accreting matter.

To understand the effect of magnetic susceptibility on the development of MRI, we follow the linear analysis of axisymmetric perturbations carried out by \citet{Balbus91}. Consider a stationary, axisymmetric accretion disc in cylindrical coordinates $(R,\phi,z)$ in which the angular velocity only depends on the radial distance, i.e. $\Omega=\Omega(R)$. On top of this equilibrium solution, we overlay a magnetic field $\mathbf{B}=B_\phi(R,z)\hat{\phi}+B_z(R)\hat{z}$, weak enough so that it does not affect the initial hydrodynamical state. For simplicity, we are going to do this analysis in the Newtonian limit, where the mass conservation, Euler, and induction equations can be written as,
\begin{align}
&\frac{d \ln \rho}{dt}+\nabla\cdot\mathbf{v}=0, \label{mass_newtonian}\\
&\frac{d\mathbf{v}}{dt}+\frac{1}{\rho}\nabla\left[p+(1-2\chi)\frac{B^{2}}{8\pi}\right]-\frac{1-\chi}{4\pi\rho}(\mathbf{B}\cdot\nabla)\mathbf{B}+\nabla\Phi=0,\label{euler_newtonian}\\
&\frac{\partial\mathbf{B}}{\partial t}-\nabla\times(\mathbf{v}\times\mathbf{B})=0\label{induction_newtonian},
\end{align}
where $\mathbf{v}$ and $\mathbf{B}$ are the spatial velocity and the spatial magnetic field, respectively, while $\Phi$ is the gravitational potential of the central object.

We apply a perturbation to the initial state of the form $\delta q=\delta q_{_0}\exp[i(k_RR+k_zz-\omega t)]$, where $\delta q_{_0}$ is a constant amplitude, $k_R$ and $k_z$ are the radial and vertical wave numbers, respectively, $\omega$ is the oscillation frequency, and $q$ stands for any fluid variable. Adding this perturbation to eqs. (\ref{mass_newtonian})-(\ref{induction_newtonian}), neglecting second order or higher terms in $\delta q$, and assuming the Boussinesq approximation, we can obtain the following linearized system of equations for the disturbances,
\begin{align}
&k_{_R}\delta v_{_R}+k_{_z}\delta v_{_z}=0 ~, \label{lin1}\\
&i\omega\delta v_{_R}-\frac{ik_{_R}}{\rho}\delta p+\frac{\delta\rho}{\rho^2}\frac{\partial p}{\partial R}+2\Omega\delta v_{_\phi}-\nonumber\\
&-(1-2\chi)\frac{ik_{_R}}{4\pi \rho}(B_{_\phi} \delta B_{_\phi}-B_{_z} \delta B_{_z})+(1-\chi)\frac{ik_{_z}}{4\pi\rho}B_{_z}\delta B_{_R}=0 ~, \label{lin2}\\
&i\omega\delta v_{_z}-\frac{ik_{_z}}{\rho}\delta p+\frac{\delta\rho}{\rho^2}\frac{\partial p}{\partial z}-\nonumber\\
&-(1-2\chi)\frac{ik_{_z}}{4\pi \rho}(B_{_\phi} \delta B_{_\phi}-B_{_z} \delta B_{_z})+(1-\chi)\frac{ik_{_z}}{4\pi\rho}B_{_z}\delta B_{_z}=0 ~, \label{lin3}\\
&i\omega\delta v_{_\phi}-\frac{\kappa^{2}}{2\Omega}\delta v_{_R}+(1-\chi)\frac{ik_{_z}}{4\pi\rho}B_{_z}\delta B_{_\phi}=0 ~, \label{lin4}\\
&i\omega\delta B_{_R}+ik_{_z}B_{_z}\delta v_{_R}=0 ~, \label{lin5}\\
&i\omega\delta B_{_z}+ik_{_z}B_{_z}\delta v_{_z}=0 ~, \label{lin6}\\
\end{align}
and
\begin{align}
&i\omega\delta B_{_\phi}+\frac{d\Omega}{d\ln R}\delta B_{_R}+ik_{_z}B_{_z}\delta v_{_\phi}=0 ~, \label{lin7}
\end{align}
where $\kappa$ is the epicyclic frequency,
\begin{equation}
\kappa^{2}=\frac{2\Omega}{R}\frac{d(R^2\Omega)}{dR} ~.
\label{epicyclic}
\end{equation}
In addition, as in \citet{Balbus91}, we use the entropy for adiabatic perturbations in the Boussinesq approximation,
\begin{equation}
i\omega\frac{5}{3}\frac{\delta\rho}{\rho}+\delta v_{_z}\frac{\partial\ln \left(p\rho^{-5/3}\right)}{\partial z}+\delta v_{_R}\frac{\partial\ln \left(p\rho^{-5/3}\right)}{\partial R}=0 ~,
\label{adiabaticity}
\end{equation}
in order to close the system of equations.

We can proceed to find the dispersion relation in the usual way. Doing so, we find that the unstable modes are the complex roots with positive imaginary parts of the equation,
\begin{equation}
\frac{k_{_R}^2+k_{_z}^2}{k_{_z}^2}\tilde{\omega}^4-\left[\kappa^2+\left(\frac{k_{_R}}{k_{_z}}N_{_z}-N_{_R}\right)^2\right]\tilde{\omega}^2-4\Omega^2k_{_z}^2\tilde{v}_{A_z}^2=0 ~,
\label{dispersion}
\end{equation} 
where $\tilde{\omega}^2=\omega^2-k_{_z}^2\tilde{v}_{A_z}^2$ and $\tilde{v}_{A_z}^2=(1-\chi)v_{A_z}^2$, $v_{A_z}=B_{_z}/\sqrt{4\pi\rho}$ being the vertical Alfv\'en velocity. The quantities $N_{_R}$ and $N_{_z}$ are the buoyancy frequencies in the corresponding directions. They are given by the expressions,
\begin{equation}
N_{_R}^2=-\frac{3}{5\rho}\frac{\partial p}{\partial R}\frac{\partial\ln \left(p\rho^{-5/3}\right)}{\partial R} ~,\hspace{4mm}N_{_z}^2=-\frac{3}{5\rho}\frac{\partial p}{\partial z}\frac{\partial\ln \left(p\rho^{-5/3}\right)}{\partial z} ~,
\label{buoyancy}
\end{equation}
and the total buoyancy frequency is $N^2=N_{_R}^2+N_{_z}^2$. We note that the magnetic susceptibility of the material can be included in the analysis as a simple modification of the Alfv\'en velocity, in which case the stability criteria are not directly modified by $\chi_m$, since they only depend on the hydrodynamical properties. Nevertheless, the magnetic polarization in the disc, parametrized by $\chi_m$, does modify the critical wavenumber, 
\begin{align}
k_{_{z,\text{crit}}}^2=&\frac{1+\chi_m}{2v_{A_z}^2}\left\{\left[\left(N^2+\frac{d\Omega^2}{d\ln R}\right)^2-4N_{_z}^2\frac{d\Omega^2}{d\ln R}\right]-\right.\nonumber\\
&\left.-\left(N^2+\frac{d\Omega^2}{d\ln R}\right)\right\} ~,
\label{k_crit}
\end{align}
which defines the minimum wavelength of the unstable modes. Based on equation (\ref{k_crit}), we can state that, in a paramagnetic ($\chi_m > 0$) accretion disc, the critical wavelength, $\lambda_{\text{crit,p}}$, is shorter than the critical wavelength in a diamagnetic disc ($\chi_m < 0$) , $\lambda_{\text{crit,d}}$, i.e.,
\begin{equation}
\lambda_{\text{crit,p}}<\lambda_{\text{crit,0}}<\lambda_{\text{crit,d}} ~,
\label{first1}
\end{equation}
with $\lambda_{\text{crit,0}}$ being the critical wavelength in a disc with $\chi_m=0$.

\begin{figure}
\centering
\includegraphics[width=1\columnwidth]{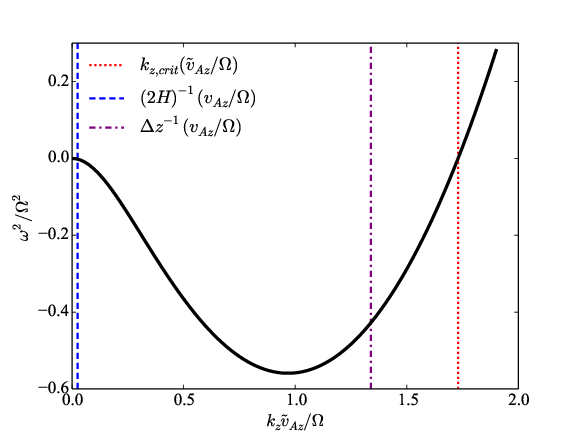}
\caption{Plot of $\omega^2/\Omega^2$ through region of instability ($\omega^2 < 0$) for an isothermal, thin, Keplerian disc. The vertical lines represent relevant scales for our MRI analysis: red, dotted is the critical wavelength; blue, dashed is the inverse of the disc thickness; and purple, dot-dashed is the inverse of the numerical resolution. The last two correspond to the start of the $\chi_m = 0$ simulation.}
\label{fig:unstable}
\end{figure}

Another important quantity that could be modified by the magnetic susceptibility is the wavevector of the fastest growing mode. To illustrate this relation, we use the same example problem as presented in \citet{Balbus91}: an isothermal, thin, Keplerian disc background \citep{Pringle81}, and the perturbations following an adiabatic evolution. In this case, it is possible to compute $N_{_R}$ and $N_{_z}$ explicitly and solve for $\omega^2$. In Fig. \ref{fig:unstable}, we plot $\omega^2/\Omega^2$ (black, solid curve) as a function of $k_{_z}\tilde{v}_{A_z}/\Omega$, for $k_{_R}=0$ corresponding to the maximum instability. We see that the most unstable mode (most negative value) is located at $k_{_z}\tilde{v}_{A_z}/\Omega\approx1$, with the corresponding wavelength of the fastest growing mode being
\begin{equation}
\lambda_{_{\text{MRI}}}\approx\frac{v_{A_z}}{\Omega\sqrt{1+\chi_m}} ~.
\label{lambda_MRI}
\end{equation}
Since $\lambda_{_{\text{MRI,p}}}<\lambda_{_{\text{MRI,d}}}$, we expect to see smaller turbulent structures in paramagnetic discs than in diamagnetic ones during the growth phase of the MRI, all other things being equal. Fig. \ref{fig:unstable} also shows the critical wavenumber, $k_{z, \mathrm{crit}}$ (red, dotted line). This, too, appears to have a fixed value when normalized by $\Omega/\tilde{v}_{A_z}$, as in the figure, but would correspond to different physical length scales for paramagnetic and diamagnetic discs (because of  the inherent dependence on $\chi_m$ in $\tilde{v}_{A_z}$).

The saturated state of the MRI is usually reached when $\lambda_{_{\text{MRI}}}\approx 2H$ \citep{Beckwith11}, $H$ being a characteristic length scale of the system, in this case the half-height of the disc. From this and equation (\ref{lambda_MRI}), we find that the magnetization parameter in the vertical direction, $\beta_{z,\text{sat}}=2p/B_{z}^2$, should scale with magnetic susceptibility according to
\begin{equation}
\beta_{z,\mathrm{sat}}=\frac{p}{8\pi H^2\Omega^2\rho}(1+\chi_m)^{-1} ~,
\label{beta_z_scale}
\end{equation} 
such that $\beta_{z,\mathrm{sat}}$ should be smaller in paramagnetic discs than in diamagnetic ones, again all other things being equal, i.e.
\begin{equation}
\beta_{\text{sat,p}}<\beta_{\text{sat,0}}<\beta_{\text{sat,d}} ~,
\label{beta_sat}
\end{equation}
where all quantities are calculated in the saturated state, with $\beta_{\text{sat,0}}$ the magnetization parameter for a disc with $\chi_m=0$.

\section{Magnetically polarized disc simulations}
\label{sec:simulations}

To test our predictions for the impact of magnetic susceptibility on black hole accretion discs, we use the {\em Cosmos}++ computational astrophysics code \citep{Anninos05}, coupled with its Constrained Transport method \citep{Fragile12}, to solve the GRMHD equations including magnetic polarization \citep{Pimentel20181}. The initial configuration used in these simulations mimics the SANE model from the EHT code comparison project \citep{Porth19}. In this model, an initial hydrodynamic torus, supported by pressure and centrifugal forces \citep{Fishbone76}, orbits around a rotating black hole with spin $a/M=0.9375$. The initial density distribution has a maximum at $r(\rho_{\text{max}})=12 r_g$ and truncates at $r_{\text{in}}=6 r_g$, where $r_g=GM/c^2$ is the gravitational radius. A weak ($\beta=p_\mathrm{max}/p_{m,\mathrm{max}}=100$, where $p_m = b^\mu b_\mu/2$), purely poloidal magnetic field with a vector potential of the form $A_{\phi}\propto\mathrm{max}(\rho/\rho_{\mathrm{max}}-0.2,0)$ is superimposed on top of this initial state to seed the MRI. This field can also be viewed as a possible cause of magnetic polarization within the disc material. We consider three different base simulations: a paramagnetic one with $\chi_m=0.\bar{3}$, a diamagnetic one with $\chi_m=-0.2$, and a non-polarized one ($\chi_m=0$) that serves as our control. The non-zero values of $\chi_m$ were chosen to be large enough to have a significant impact in the simulations and to yield similar magnitudes for $\chi$ in both the paramagnetic and diamagnetic cases. 

The initial tori are mapped onto a three-dimensional grid of $36\times 24\times 24$ zones with two or three levels of refinement applied over the whole grid, except for regions close to the pole at small radii ($\vert \theta - \pi/2 \vert \gtrsim 1.4$ and $r \lesssim 15\,GM/c^2$) for an equivalent single-level grid resolution of $144\times 96\times 96$ or $288\times 192\times 192$ zones. These resolutions are comparable to the test simulations presented in \citet{Porth19}. At our base resolution, each simulation is run to a time of  $10,000~GM/c^3$, while the higher resolution runs are stopped at $2,000~GM/c^3$.

For each grid, the radial coordinate is spaced logarithmically according to $x_1=1+\ln(r/r_g)$, between the boundaries $r_{\text{min}}=1.26r_g$ and $r_{\text{max}}=354r_g$, with outflow conditions applied at each boundary. The $\theta$ domain is covered with a grid in which  $\theta=\pi x_2+0.3\sin(2\pi x_2)$, where the coordinate $x_2$ is spaced uniformly between 0 and 1, covering from pole to pole. Ghost zones along the poles are filled with information from their corresponding real zones across the poles. The $\phi -$direction is covered with an uniformly distributed grid over the range $0\leq\phi < 2\pi$, with periodic conditions at the boundaries. These choices give proper distances of \{$0.51r_g$, $0.16r_g$, $0.79r_g$\} and \{$0.25r_g$, $0.08r_g$, $0.39r_g$\} between grid cells near the initial density maximum of the torus for the low- and high-resolution simulations, respectively. Fig. \ref{fig:unstable} shows the vertical grid resolution for our (low-resolution) $\chi_m = 0$ simulation (purple, dot-dashed line) compared to the fastest growing and critical modes of the MRI. The blue, dashed line corresponds to the disc thickness, and defines the longest possible unstable (vertical) wavelength (lowest mode) within the disc. All the possible unstable modes are, therefore, confined to the interval between the blue and red lines, while our numerical simulations are effectively confined to modes below the purple line (associated with $\Delta z$).

\subsection{MRI resolution}

\begin{figure}
\centering
\includegraphics[width=1\columnwidth]{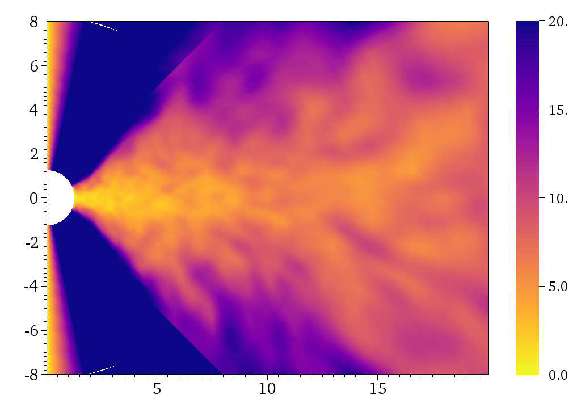}
\includegraphics[width=1\columnwidth]{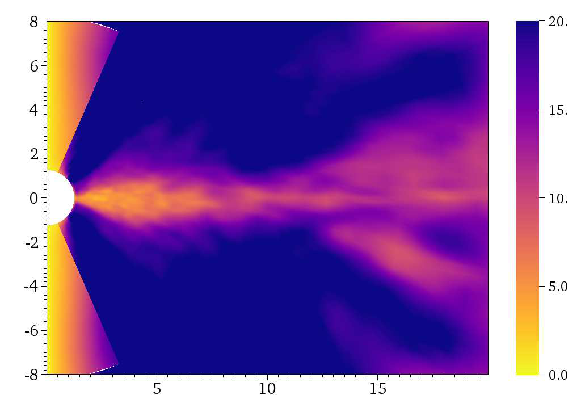}
\caption{Spatial variance of the $Q_\theta$ (top) and $Q_\phi$ (bottom) parameters in the $r$-$\theta$ plane for the $\chi_m = 0$ simulation. Data are time-averaged over the period of maximum growth of the MRI, $500 \le t/(GM/c^3) \le 2000$.}
\label{fig:Q}
\end{figure}

For a given number of refinement levels, each of our numerical simulations uses the same computational grid. However, because the MRI wavelength depends on the magnetic susceptibility [eq. (\ref{lambda_MRI})], each simulation has a slightly different effective resolution with respect to the MRI. A standard way to quantify the MRI resolution is through the $Q$ parameter
\begin{equation}
Q_i = \frac{\lambda_{\mathrm{MRI},i}}{\Delta x_i} ~,
\end{equation}
where $i$ is an index representing each coordinate direction and $\Delta x_i$ is the proper zone length in that direction. Previous resolution studies \citep[e.g.,][]{Hawley11,Hawley13} have suggested that global simulations require $Q_\theta \gtrsim 10$ and $Q_\phi \gtrsim 20$ to see well converged development of the MRI. Our $\chi_m = 0$ simulation is, on the whole, marginally consistent with these requirements, as shown in Fig. \ref{fig:Q}. Initially, since all three low-resolution simulations start with the same grid and magnetic field configuration, the $\chi_m = -0.2$ case effectively has a 12\% better $Q$ value (in all directions), because of its longer MRI wavelength. Conversely, the $\chi_m = 0.\bar{3}$ case initially has an approximately 13\% worse $Q$, because of its shorter MRI wavelength. However, as the simulations evolve and the magnetization of the discs change, the MRI wavelengths and $Q$ parameters also vary. The strong magnetization of the $\chi_m = 0.\bar{3}$ case ultimately leads to it having the highest $Q$ parameter, particularly in the azimuthal direction.

\section{Simulation Results}
\label{sec:results}

We now analyze our simulations to assess what effect magnetic susceptibility has on the development and saturation of the MRI and evolution of the accretion disc. 

\subsection{Characteristic spatial scales}

\begin{figure}
\centering
\includegraphics[width=1\columnwidth]{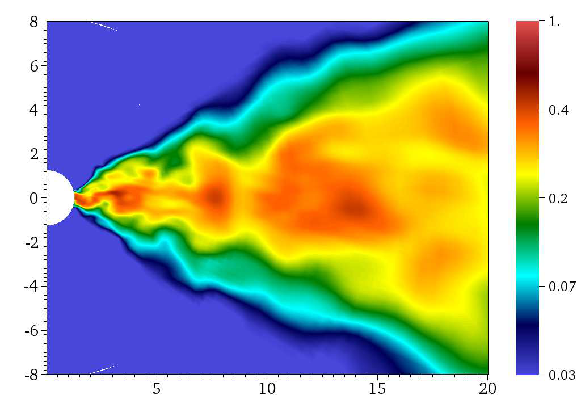}
\includegraphics[width=1\columnwidth]{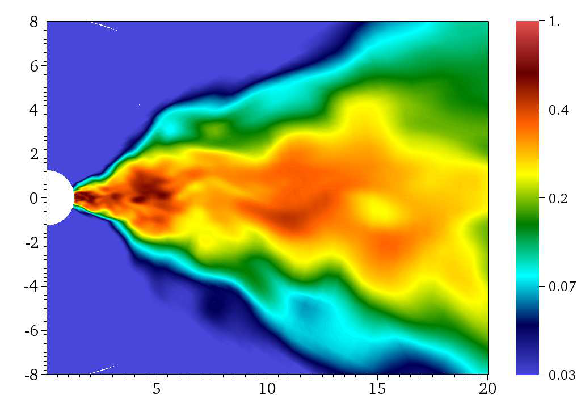}
\includegraphics[width=1\columnwidth]{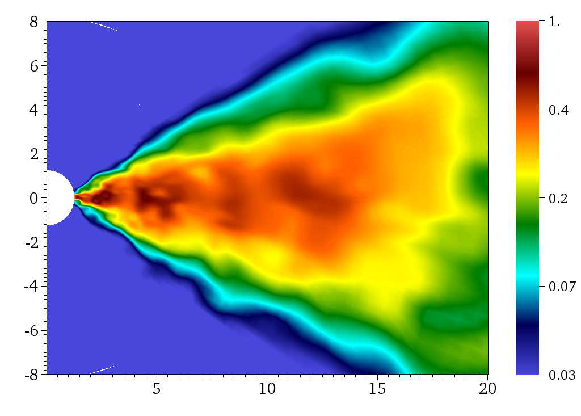}
\caption{Poloidal slices showing pseudo-color plots of the density, $\rho$, for the low-resolution paramagnetic ($\chi_m = 0.\bar{3}$; top), non-polarized ($\chi_m = 0.0$; middle), and diamagnetic ($\chi_m = -0.2$; bottom) simulations at $t=2000~GM/c^3$. Data have been spatially smoothed for this plot.}
\label{fig:rho}
\end{figure}

The first prediction we made for magnetically polarized discs was that we expect to see smaller critical and most-unstable wavelengths for paramagnetic than diamagnetic discs. However, this only applies during the linear growth phase of the MRI and presumes that all other quantities (besides $\chi_m$) remain fixed. It is more practical for global simulations and frankly of more interest to look at the non-linear saturated state of the discs. Fig. \ref{fig:rho}, which shows poloidal slices of density, provides a crude first look at the turbulent structure within each disc. 

To be more quantitative, we follow the method of \citet{Beckwith11, Hogg16} to find the correlation lengths in each direction over time for each simulation and use that as a measure of the characteristic length scales of the turbulence. The first step in this procedure is to define an analysis region, which we choose to be $r \in [5r_g, 15r_g]$, $\theta \in [\pi/2-1/3, \pi/2+1/3]$, and $\phi \in [0, 2\pi]$. This covers approximately one scale height above and below the midplane and a radial range far enough out to avoid the innermost stable circular orbit region, yet close enough in to capture the growth and saturation of the MRI over the timescales of these simulations. The number of zones in this region works out to be 28, 38, and 96 in the $r$, $\theta$, and $\phi$ directions, respectively, for the low-resolution simulations and twice that in each direction for the high-resolution simulations. 

Next, we subtract off a time average of the background field, leaving only the residual fluctuations. The time average is taken over a span of eleven consecutive data dumps, five before and five after the time, $t$, of interest. The resulting calculation for a generic scalar field $F$ looks like
\begin{equation}
F_{\mathrm{sub}}(x_1, x_2, \phi, t) = F(x_1, x_2, \phi, t) - \frac{1}{\Delta T} \int_{t - 0.5\Delta T}^{t + 0.5\Delta T} F(x_1, x_2, \phi, t) dt ~, 
\end{equation}
where $\Delta T = 500~GM/c^3$ is the averaging window. We tried this procedure for $\rho$, $B^r$, $B^\theta$, and $B^\phi$. As we show below, we find almost no systematic variation in the magnetic field variables from one simulation to the next, and only a marginal dependence in the characteristic scales of density on $\chi_m$. This behavior doesn't agree with our prediction from equation (\ref{lambda_MRI}), since we were expecting to obtain smaller structures in paramagnetic disks than in diamagnetic ones.

The next step is to take the three-dimensional Fourier transform 
\begin{eqnarray}
\mathcal{F}(k_{x_1}, k_{x_2}, k_\phi, t) & = & \iiint F_{\mathrm{sub}}(x_1, x_2, \phi, t) \\
 & & \times e^{i\left(k_{x_1} \Delta x_1 + k_{x_2} \Delta x_2+ k_\phi \Delta \phi\right)} dx_1 dx_2 d\phi ~ \nonumber
\label{Fourier}
\end{eqnarray}
of the density residuals, where $\Delta x_1$, $\Delta x_2$, and $\Delta \phi$ are the coordinate differences between the point \{$x_1$, $x_2$, $\phi$\} and the center of our analysis region, and $k_{x_1} = 2 \pi i / x_{1,\mathrm{box}}$, $k_{x_2} = 2 \pi j / x_{2,\mathrm{box}}$, and $k_\phi = 2 \pi k / \phi_\mathrm{box}$ give the set of wavenumbers. The indices $i$, $j$, and $k$ range over the number of zones in a given direction of the analysis region, and $x_{1,\mathrm{box}}$, $x_{2,\mathrm{box}}$, and $\phi_\mathrm{box}$ are the coordinate lengths in each direction of our analysis region. This allows the widest range of possible wavenumbers to be considered.

\begin{figure*}
\centering
\includegraphics[width=\textwidth]{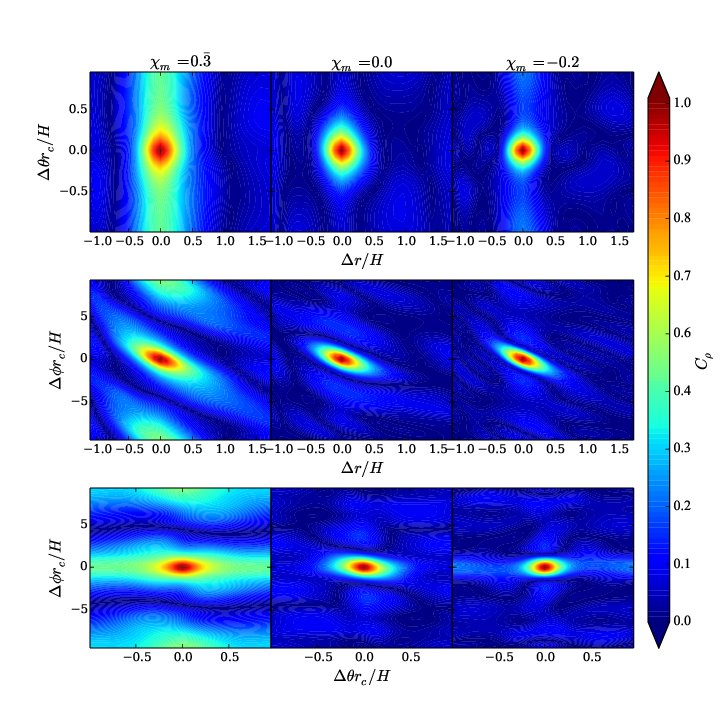}
\caption{Slices of the autocorrelation function, $C_\rho$, through the $\Delta \phi = 0$, $\Delta x_2 = 0$, and $\Delta x_1 = 0$ planes for each of the three low-resolution simulations at $t=5000~GM/c^3$. Each simulation is normalized independently.}
\label{fig:corr}
\end{figure*}

To get the autocorrelation function, we then perform an inverse Fourier transform on the power spectrum $|\mathcal{F}(k_{x_1}, k_{x_2}, k_\phi, t)|^2$ as
\begin{eqnarray}
C_\mathcal{F}(\Delta x_1, \Delta x_2, \Delta \phi, t) & = & \iiint |\mathcal{F}(k_{x_1}, k_{x_2}, k_\phi, t)|^2  \\
 & & \times e^{i\left(k_{x_1} \Delta x_1 + k_{x_2} \Delta x_2 + k_\phi \Delta \phi\right)} dk_{x_1} dk_{x_2} dk_\phi ~.  \nonumber
\label{auto_correlation}
\end{eqnarray}
Fig. \ref{fig:corr} shows slices of the resulting autocorrelation, $C_\rho$, through the $\Delta \phi = 0$, $\Delta x_2 = 0$, and $\Delta x_1 = 0$ planes for the three low-resolution simulations. The scale in each plot is normalized by the value of the autocorrelation function at the center of the box for each simulation and time step. We choose a particularly illustrative time step ($t=5000~GM/c^3$) that shows the typical ranking of the correlation lengths, with the paramagnetic case ($\chi_m = 0.\bar{3}$) exhibiting the largest turbulent structures, the diamagnetic case ($\chi_m = -0.2$) the smallest ones, and the $\chi_m = 0$ case in between. For this particular time slice, the turbulent structures (or equivalently the correlation length) are barely resolved (only spans a few zones) in the diamagnetic case, while in the paramagnetic case, they extend further than one scale height in the $\theta$ direction.

\begin{figure}
\centering
\includegraphics[width=\columnwidth]{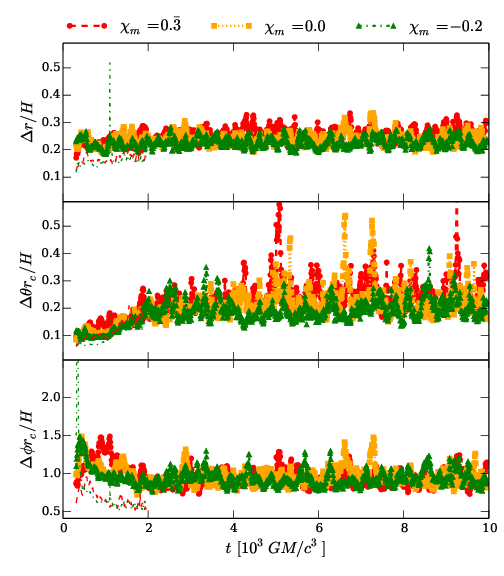}
\caption{Instantaneous values of the characteristic length scales of the turbulence, defined as where $C_\rho = 0.5$ along the three coordinate axes, for each of the five simulations. Thicker lines with symbols are used for the three low-resolution simulations, while thin lines are used for the two high-resolution ones.}
\label{fig:lengths}
\end{figure}

To support our claim that Fig. \ref{fig:corr} is reasonably representative of the results of our simulations, in Fig. \ref{fig:lengths} we track the characteristic length scales, defined as the value of $\Delta x_i$ where $C_\rho = 0.5$ along the three coordinate axes, for all five simulations as a function of time. Note that we choose $r_c = 10r_g$ and $H=10/3 r_g$. Also, Table \ref{tab:lengths} reports the means and standard deviations that come from Fig. \ref{fig:lengths} and similar plots of $B^r$, $B^\theta$, and $B^\phi$. While there is considerable variability about the mean, we only find marginal evidence in the $\theta$-direction for the paramagnetic simulations to exhibit the largest structures and the diamagnetic simulations the smallest ones. No pattern clearly emerges in the $r$- nor $\phi$-directions. Additionally, we find that $r_c\Delta\theta$ tends to be slightly smaller than $\Delta r$, while both are more than a factor of four smaller than $r_c\Delta\phi$. This behavior is consistent across all five simulations and therefore is roughly independent of the magnetic susceptibility. 

\begin{table}
\begin{center}
\begin{tabular}{c c c c c}
\hline
 & {Sim} & {$\Delta r/H$} & {$\Delta \theta r_c/H$} & {$\Delta \phi r_c/H$} \\
\hline
\multirow{3}{2em}{$\rho$} & $\chi_m = 0.\bar{3}$ & $0.25\pm0.02$ & $0.23\pm0.07$ & $0.96\pm0.13$ \\
& $\chi_m = 0.0$ & $0.24\pm0.02$ & $0.21\pm0.06$ & $0.97\pm0.12$ \\
& $\chi_m = -0.2$ & $0.23\pm0.02$ & $0.18\pm0.05$ & $0.95\pm0.11$ \\
 \hline
\multirow{3}{2em}{$B^r$} & $\chi_m = 0.\bar{3}$ & $0.22\pm0.02$ & $0.10\pm0.01$ & $0.67\pm0.14$ \\
& $\chi_m = 0.0$ & $0.21\pm0.02$ & $0.10\pm0.01$ & $0.69\pm0.11$ \\
& $\chi_m = -0.2$ & $0.21\pm0.02$ & $0.10\pm0.01$ & $0.69\pm0.10$ \\
 \hline
\multirow{3}{2em}{$B^\theta$} & $\chi_m = 0.\bar{3}$ & $0.16\pm0.01$ & $0.16\pm0.02$ & $0.65\pm0.11$ \\
& $\chi_m = 0.0$ & $0.16\pm0.01$ & $0.16\pm0.02$ & $0.69\pm0.10$ \\
& $\chi_m = -0.2$ & $0.16\pm0.01$ & $0.15\pm0.01$ & $0.70\pm0.08$ \\
 \hline
\multirow{3}{2em}{$B^\phi$} & $\chi_m = 0.\bar{3}$ & $0.19\pm0.01$ & $0.13\pm0.01$ & $0.89\pm0.17$ \\
& $\chi_m = 0.0$ & $0.18\pm0.02$ & $0.13\pm0.02$ & $0.97\pm0.20$ \\
& $\chi_m = -0.2$ & $0.18\pm0.02$ & $0.12\pm0.01$ & $1.05\pm1.28$ \\
 \hline
\end{tabular}
\end{center}
\caption{Mean correlation lengths of $\rho$, $B^r$, $B^\theta$, and $B^\phi$, with standard deviations,
for the three low-resolution simulations that have reached a saturated state.}
\label{tab:lengths}
\end{table}

Comparing our results with previous measurements of the correlation length in non-polarized disc simulations, we find our results to be qualitatively and quantitatively similar. \citet{Hogg16} found a nearly constant $\Delta r/H$ value of approximately 0.32 for $\rho$, which is similar to what we find for the $\chi_m = 0$ simulation in the top panel of Fig. \ref{fig:lengths} and Table \ref{tab:lengths}. The middle column of Fig. \ref{fig:corr} also agrees well with Fig. 8 of \citet{Hogg16}. These results, though, may depend upon resolution, as suggested by the trends we see in our data. Especially, in the $r$ and $\phi$ directions, the correlation lengths dropped appreciably when we increased our resolution, as seen in Fig. \ref{fig:lengths}.

\subsection{Magnetization parameter}

\begin{figure}
\centering
\includegraphics[width=1\columnwidth]{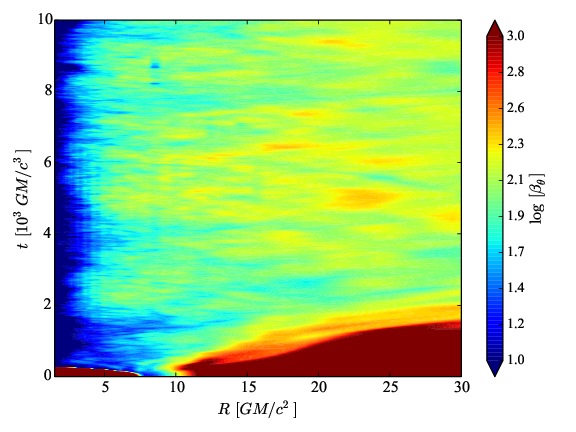}
\includegraphics[width=1\columnwidth]{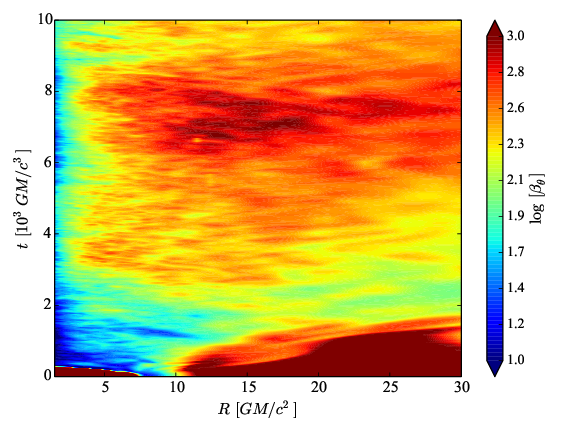}
\includegraphics[width=1\columnwidth]{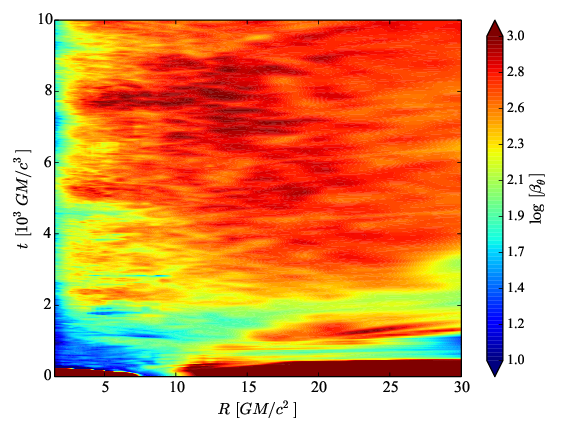}
\caption{Spacetime diagrams of the magnetization parameter, $\langle \beta_\theta \rangle_\rho$, for the low-resolution paramagnetic ($\chi_m = 0.\bar{3}$; top), non-polarized ($\chi_m = 0.0$; middle), and diamagnetic ($\chi_m = -0.2$; bottom) simulations.}
\label{fig:magnetization}
\end{figure}

The next prediction we made regarding magnetically polarized discs was that the magnetization parameter, $\beta_{z,\mathrm{sat}}$, should be smaller for paramagnetic discs than for diamagnetic ones. Fig. \ref{fig:magnetization}, which shows density-weighted shell averages of $\beta_\theta = 2p/(g_{\theta\theta} B^\theta B^\theta)$, which is a convenient proxy for $\beta_z$, for the three low-resolution simulations, supports this assertion. These averages are calculated as
\begin{equation}
\left< \beta_\theta \right>_\rho = \frac{\iint \beta_\theta \sqrt{-g} \rho\left( x_1, x_2, \phi \right) dx_2 d\phi}{\iint \sqrt{-g} \rho \left( x_1, x_2, \phi \right) dx_2 d\phi}.
\end{equation}
The top plot, corresponding to the paramagnetic ($\chi_m = 0.\bar{3}$) disc, shows that it becomes more strongly magnetized ($\beta_\theta \lesssim 100$) than the $\chi_m=0$ case, shown in the middle plot, and the $\chi_m=-0.2$ case in the bottom plot, which saturates at $\beta_\theta \gtrsim 1000$. We conclude then that diamagnetism indeed increases the magnetization parameter inside an accretion disc, while paramagnetism reduces it, and therefore increases the role of the magnetic field in the disc. The observed differences are somewhat larger than the $1/(1+\chi_m)$ factor one might naively expect from equation (\ref{beta_sat}). These differences probably result from the fact that the saturation condition we use in our analysis is based only on a geometrical argument, but the non-linear saturation of the MRI is also influenced by the dissipation coefficients \citep{lesur2007impact, pessah2009saturation}. Therefore, it is necessary to explicitly include viscous and resistive effects in the analysis to carry out a more appropriate quantitative study of the MRI saturation in magnetically polarized media.

The higher level of magnetization in the paramagnetic simulation is potentially important for a number of reasons. First, the turbulence driven by the MRI strongly depends on the vertical magnetic field in such a way that the efficiency of the angular momentum transport in the radial direction and the mass outflow rate increase for lower values of $\beta_\theta$ \citep{suzuki2010protoplanetary,bai2013local}. Hence, the magnetic susceptibility could improve the efficiency of the accretion process driven by the MRI. Strong magnetic fields may also provide a means to solve the thermal and viscous instability problems associated with geometrically thin, Shakura-Sunyaev-type discs \citep{Begelman07,Oda09,Sadowski16}. Furthermore, strong, vertical magnetic fields are useful for driving winds \citep{Blandford82} and powering jets \citep{Blandford77,McKinney06}, both features commonly associated with accreting black holes. Additionally, high magnetic pressure may allow for higher mass accretion rates in self-gravitating AGN discs \citep{Begelman07}.

\subsection{Effective viscosity \& mass accretion}

\begin{figure}
\centering
\includegraphics[width=1\columnwidth]{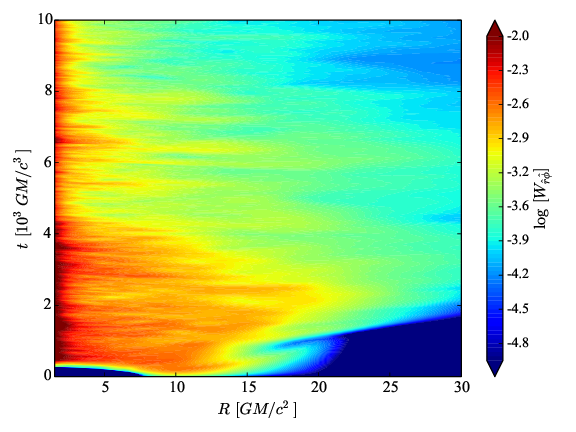}
\includegraphics[width=1\columnwidth]{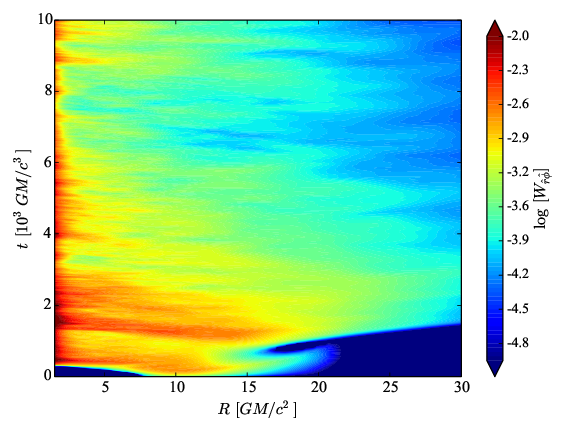}
\includegraphics[width=1\columnwidth]{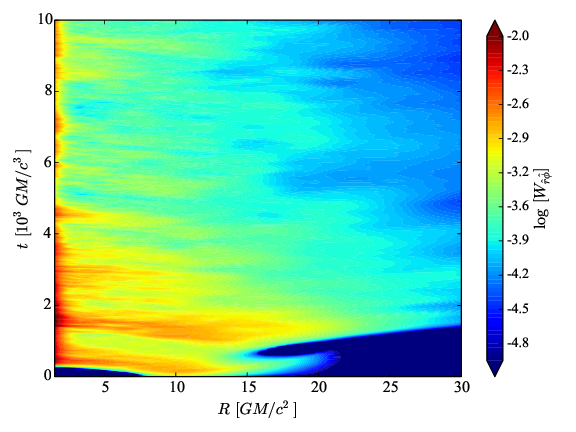}
\caption{Spacetime diagrams of the vertically integrated shear stress, $W_{\hat{r}\hat{\phi}}$, for the low-resolution paramagnetic ($\chi_m = 0.\bar{3}$; top), non-polarized ($\chi_m = 0.0$; middle), and diamagnetic ($\chi_m = -0.2$; bottom) simulations.}
\label{fig:alpha}
\end{figure}

Not surprisingly then, changes to the magnetic polarization and magnetization parameter translate into changes in the integrated stress and mass accretion histories of each simulation. For example, Fig. \ref{fig:alpha} shows the vertically integrated and azimuthally averaged fluid-frame shear stress, 
\begin{equation}
W_{\hat{r}\hat{\phi}} = \frac{\int_0^{2\pi}\int_{-H}^{+H} e^\mu_{(r)} e^\nu_{(\phi)}T_{\mu\nu}\sqrt{-g}\mathrm{d}r\mathrm{d}\theta\mathrm{d}\phi}{\int_0^{2\pi} r \mathrm{d}r\mathrm{d}\phi} ~,
\end{equation} 
where $e^\mu_{(\nu)}$ are the set of basis vectors in the fluid rest frame. The discs in our simulations all start with small values of stress. Nevertheless, it grows rapidly in the region $r\lesssim 15 r_g$, peaking at most radii around a time of $t\lesssim 2000~GM/c^3$. After that, $W_{\hat{r}\hat{\phi}}$ gradually declines for most of the rest of each simulation. The key result, in terms of our present study, is that the magnitude of the stress depends on the magnetic susceptibility, with the paramagnetic simulation showing the highest stresses and the diamagnetic showing the lowest. This ordering is consistent with the magnetizations measured in the previous section.

\begin{figure}
\centering
\includegraphics[width=1\columnwidth]{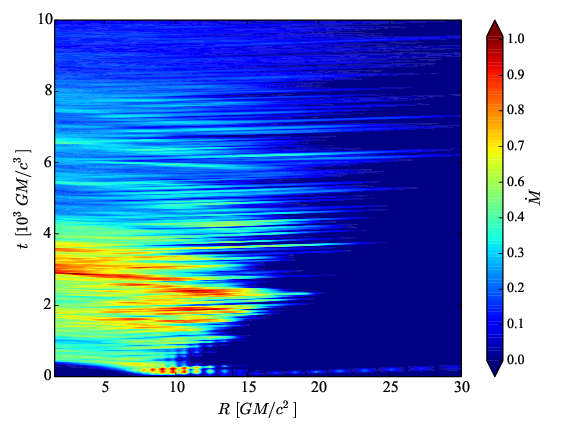}
\includegraphics[width=1\columnwidth]{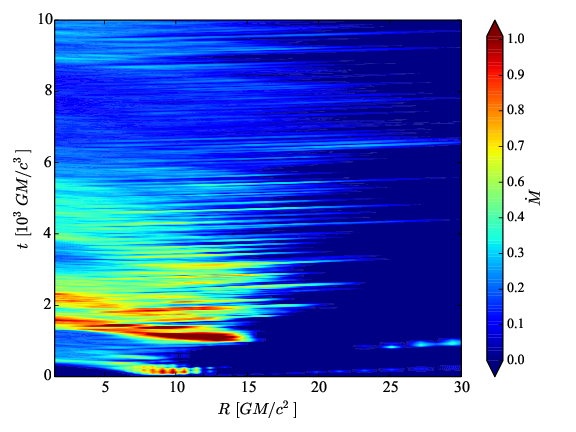}
\includegraphics[width=1\columnwidth]{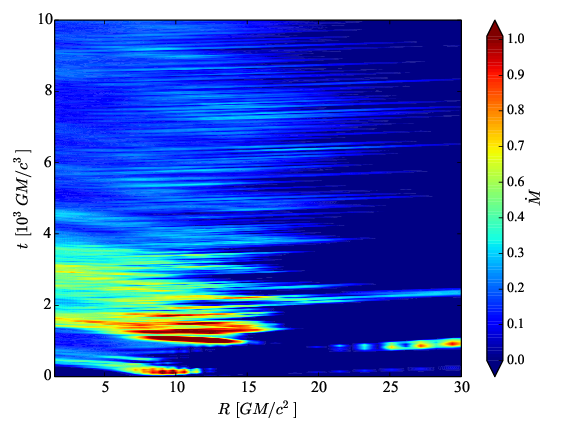}
\caption{Spacetime diagrams of the mass accretion rate, $\dot{M}$, in arbitrary units, defined to be positive for material flowing toward the black hole, for the low-resolution paramagnetic ($\chi_m = 0.\bar{3}$; top), non-polarized ($\chi_m = 0.0$; middle), and diamagnetic ($\chi_m = -0.2$; bottom) simulations.}
\label{fig:mdot}
\end{figure}

In steady-state, geometrically thin discs, the mass accretion rate, $\dot{M}$, is directly proportional to the vertically integrated stress \citep{Shakura73,Novikov73}. Although our simulations are of relatively thick, finite tori, we, nevertheless, would expect the mass accretion rate to depend upon the stress. Furthermore, since the stress is dependent upon the magnetic susceptibility, we would also expect the mass accretion rate to depend on $\chi_m$ (all other elements of the stress being equal). We find some evidence for a dependence of $\dot{M}$ on $\chi_m$ in Fig. \ref{fig:mdot}, which shows spacetime diagrams of the mass accretion rate for each low-resolution simulation, though the correspondence is weak. Notably, the simulations exhibit larger differences in $W_{\hat{r}\hat{\phi}}$ than they do in $\dot{M}$.

However, $\dot{M}$ is a tricky diagnostic to deal with in these tori simulations, which only have a finite mass reservoir to deal with. In the case of the $\chi_m = 0.\bar{3}$ simulation, the mass accretion rate drops at late time partly because of the loss of available matter. By the end of the low-resolution simulation, the $\chi_m = 0.\bar{3}$ case only has about 63\% of its initial mass remaining (vs. 70\% and 72\% for the $\chi_m = 0$ and -0.2 simulations, respectively).

Another issue with the finite torus setup is that conservation of momentum requires that some of the torus material gain angular momentum and consequently move outward. This outward radial motion is indicated by the deep blue colors in Fig. \ref{fig:mdot}. Over time, the stagnation point in the disk (the separation point between in-flowing and out-flowing material) moves toward larger radii, yet it still affects the overall $\dot{M}$ that can be sustained.

\section{Discussion \& Conclusions}
\label{sec:conclusions}

In this paper we analyzed the MRI in magnetically polarized plasmas by studying the growth of perturbations in a weakly magnetized accretion disc. The linear analysis of the instability, which is based on the seminal paper of \citet{Balbus91}, showed that the growth rate of the disturbances is independent of the magnetic susceptibility $\chi_m$, but the wavelengths of the critical and fastest growing modes are shorter in paramagnetic fluids than in diamagnetic ones. Additionally, by assuming that the saturation of the MRI is reached when $\lambda_{\text{MRI}}\approx 2H$, we conclude that the magnetization parameter in that state should be larger in diamagnetic discs than in paramagnetic ones, i.e. $\beta_{\text{sat,p}}<\beta_{\text{sat,0}}<\beta_{\text{sat,d}}$. The theoretical relationship between $\lambda_{\text{MRI}}$ and the magnetic susceptibility further suggests that the polarization will have an effect on turbulent processes, and therefore angular momentum transport, within accretion discs.

In order to test the predictions from our linear analysis, we carried out global simulations of magnetically polarized accretion discs with different magnetic susceptibilities. The evolution with $\chi_m=0.\bar{3}$ and $\chi_m=-0.2$ describe the accretion of a paramagnetic and a diamagnetic disc, respectively, and the simulation with $\chi_m=0$ corresponds to the usual unpolarized disc, which is useful for comparison purposes. We first analyzed the turbulence in the disc through the correlation lengths of different fluid variables and found marginal dependence on the magnetic polarization for gas density. There appears to be a tendency for the paramagnetic disc to have larger turbulent structures in the vertical direction than in the diamagnetic case. The trend is unclear in the $r$- and $\phi$-directions.

The magnetization parameter $\beta=p/p_m$ shows that the paramagnetic disc becomes more strongly magnetized than the $\chi_m=0$ case, which itself is more strongly magnetized than the diamagnetic disc. This behaviour was predicted by the linear analysis in equation (\ref{beta_sat}), though the differences are larger than the expected $1/(1+\chi_m)$ factor. The main reason for these differences probably comes from the fact that the saturation condition we used in the linear analysis was based on a purely geometrical argument, but the saturation level also depends on dissipative processes. Therefore, subsequent quantitative studies of MRI saturation should explicitly include viscous and resistive effects.

Finally, we have found that the vertically integrated shear stress depends on the magnetic susceptibility of the fluid. Even though the simulations start with a small stress, it grows rapidly (especially in regions near the event horizon) in such a way that the paramagnetic disc reaches the highest values of stress while the diamagnetic one achieves the lowest. This result is consistent with the behaviour of the vertical magnetization parameter and it is interesting because $W_{\hat{r}\hat{\phi}}$ is the component of the stress responsible for the angular momentum transport within the disc. In fact, we have also found a dependence of the mass accretion rate, $\dot{M}$, on magnetic susceptibility, though not as strong as might be expected from the behaviour of $W_{\hat{r}\hat{\phi}}(\chi_m)$. This likely has to do with the finite mass reservoir and outward migrating stagnation point that are hallmarks of these simulations.

As mentioned in the introduction, the magnetic polarization of matter is usually small when the magnetic fields are weak, so its effects on the dynamics of plasmas are often not relevant. Nevertheless, the interaction between the ambient magnetic field and the electrons bound in the atoms can grow when the energy of the magnetic dipoles is comparable with the thermal energy of the fluid. In low mass X-ray binaries, where accretion discs have temperatures in the range of $10^{6-7}$ K, magnetic fields $\gtrsim 10^{10}$ G are required for the paramagnetism of the fluid to deviate from the Curie law and possibly have non-negligible effects. Another factor that can increase the diamagnetic response is to have a high-density plasma \citep{ray2001atomic}.

One astrophysical object that satisfies these criteria is a neutron star, where the surface magnetic field can be in the range of $10^{12-15}$ G and the mass density is $\sim 10^{13}$ g cm$^{-3}$. In fact, neutron stars are considered ideal astrophysical laboratories for testing for diamagnetic phase transitions and the formation of magnetic domains \citep{blandford1982magnetic, broderick2000equation, wang2016diamagnetic}. Therefore, a neutron star tidally disrupted by a black hole represents one plausible astrophysical scenario where our analytical and numerical results may be applied. Plus, there is more work to be done to understand how large the effective magnetic susceptibility parameter is for hot ($T \gtrsim 10^5$ K), low density ($\rho < 10^2$ g cm$^{-3}$) plasmas in order to check how relevant our results are to ordinary accretion discs. We leave this for future work.

Another scenario where the magnetic polarization might be relevant to the MRI is a protoneutron star. In such a high-density system with strong magnetic fields, the MRI is considered as a promising mechanism for amplifying local magnetic fields in a rapidly rotating core to produce large-scale, ordered magnetic dipoles, and therefore to explain the formation of magnetars \citep{mosta2015large, reboul2021global}. Based on the results obtained in this work, and supported by previous studies \citep{Pimentel2019}, the polarization could modify the strength of the amplified magnetic field, with paramagnetic fluids possibly generating higher magnetic dipoles. Nevertheless, a detailed analysis of the magnetic susceptibility in protoneutron stars requires numerical simulations accounting for their specific properties \citep{obergaulinger2009semi}. 

\section*{Acknowledgements}

The authors would like to thank Omer Blaes and Drew Hogg for helpful discussions related to this work. O. M. P. wants to thanks the financial support from COLCIENCIAS under the program Becas Doctorados Nacionales 647 and Universidad Industrial de Santander. F.D.L-C were supported by the Vicerrectoría de Investigación y Extensión - Universidad Industrial de Santander, under Grant No. 2493. P.C.F. acknowledges support from National Science Foundation grant AST-1616185, AST-1907850, and PHY-1748958. This work used the Extreme Science and Engineering Discovery Environment (XSEDE), which is supported by National Science Foundation grant number ACI-1053575.

\section*{Data Availability}

The data underlying this article will be shared on reasonable request to the corresponding author.


\begin{thebibliography}{}
\makeatletter
\relax
\def\mn@urlcharsother{\let\do\@makeother \do\$\do\&\do\#\do\^\do\_\do\%\do\~}
\def\mn@doi{\begingroup\mn@urlcharsother \@ifnextchar [ {\mn@doi@}
  {\mn@doi@[]}}
\def\mn@doi@[#1]#2{\def\@tempa{#1}\ifx\@tempa\@empty \href
  {http://dx.doi.org/#2} {doi:#2}\else \href {http://dx.doi.org/#2} {#1}\fi
  \endgroup}
\def\mn@eprint#1#2{\mn@eprint@#1:#2::\@nil}
\def\mn@eprint@arXiv#1{\href {http://arxiv.org/abs/#1} {{\tt arXiv:#1}}}
\def\mn@eprint@dblp#1{\href {http://dblp.uni-trier.de/rec/bibtex/#1.xml}
  {dblp:#1}}
\def\mn@eprint@#1:#2:#3:#4\@nil{\def\@tempa {#1}\def\@tempb {#2}\def\@tempc
  {#3}\ifx \@tempc \@empty \let \@tempc \@tempb \let \@tempb \@tempa \fi \ifx
  \@tempb \@empty \def\@tempb {arXiv}\fi \@ifundefined
  {mn@eprint@\@tempb}{\@tempb:\@tempc}{\expandafter \expandafter \csname
  mn@eprint@\@tempb\endcsname \expandafter{\@tempc}}}

\bibitem[\protect\citeauthoryear{{Anninos}, {Fragile}  \&
  {Salmonson}}{{Anninos} et~al.}{2005}]{Anninos05}
{Anninos} P.,  {Fragile} P.~C.,   {Salmonson} J.~D.,  2005, \mn@doi [\apj]
  {10.1086/497294}, \href
  {https://ui.adsabs.harvard.edu/\#abs/2005ApJ...635..723A} {635, 723}

\bibitem[\protect\citeauthoryear{{Ant{\'o}n}, {Zanotti}, {Miralles},
  {Mart{\'{\i}}}, {Ib{\'a}{\~n}ez}, {Font}  \& {Pons}}{{Ant{\'o}n}
  et~al.}{2006}]{2006ApJ...637..296A}
{Ant{\'o}n} L.,  {Zanotti} O.,  {Miralles} J.~A.,  {Mart{\'{\i}}} J.~M.,
  {Ib{\'a}{\~n}ez} J.~M.,  {Font} J.~A.,   {Pons} J.~A.,  2006, \mn@doi [\apj]
  {10.1086/498238}, \href {http://adsabs.harvard.edu/abs/2006ApJ...637..296A}
  {637, 296}

\bibitem[\protect\citeauthoryear{{Bai} \& {Stone}}{{Bai} \&
  {Stone}}{2013}]{bai2013local}
{Bai} X.-N.,  {Stone} J.~M.,  2013, \mn@doi [\apj]
  {10.1088/0004-637X/767/1/30}, \href
  {https://ui.adsabs.harvard.edu/abs/2013ApJ...767...30B} {767, 30}

\bibitem[\protect\citeauthoryear{{Balbus} \& {Hawley}}{{Balbus} \&
  {Hawley}}{1991}]{Balbus91}
{Balbus} S.~A.,  {Hawley} J.~F.,  1991, \mn@doi [\apj] {10.1086/170270}, \href
  {http://adsabs.harvard.edu/abs/1991ApJ...376..214B} {376, 214}

\bibitem[\protect\citeauthoryear{{Balbus} \& {Hawley}}{{Balbus} \&
  {Hawley}}{1998}]{Balbus98}
{Balbus} S.~A.,  {Hawley} J.~F.,  1998, \mn@doi [Reviews of Modern Physics]
  {10.1103/RevModPhys.70.1}, \href
  {https://ui.adsabs.harvard.edu/abs/1998RvMP...70....1B} {70, 1}

\bibitem[\protect\citeauthoryear{Banyuls, Font, Ib{\'a}{\~n}ez, Mart{\'\i}  \&
  Miralles}{Banyuls et~al.}{1997}]{banyuls1997numerical}
Banyuls F.,  Font J.~A.,  Ib{\'a}{\~n}ez J.~M.,  Mart{\'\i} J.~M.,   Miralles
  J.~A.,  1997, The Astrophysical Journal, 476, 221

\bibitem[\protect\citeauthoryear{Beaurepaire, Scheurer, Krill  \&
  Kappler}{Beaurepaire et~al.}{2001}]{beaurepaire2001magnetism}
Beaurepaire E.,  Scheurer F.,  Krill G.,   Kappler J.-P.,  2001, Magnetism and
  synchrotron radiation.
 Vol. 34, Springer

\bibitem[\protect\citeauthoryear{{Beckwith}, {Armitage}  \& {Simon}}{{Beckwith}
  et~al.}{2011}]{Beckwith11}
{Beckwith} K.,  {Armitage} P.~J.,   {Simon} J.~B.,  2011, \mn@doi [\mnras]
  {10.1111/j.1365-2966.2011.19043.x}, \href
  {https://ui.adsabs.harvard.edu/abs/2011MNRAS.416..361B} {416, 361}

\bibitem[\protect\citeauthoryear{{Begelman} \& {Pringle}}{{Begelman} \&
  {Pringle}}{2007}]{Begelman07}
{Begelman} M.~C.,  {Pringle} J.~E.,  2007, \mn@doi [\mnras]
  {10.1111/j.1365-2966.2006.11372.x}, \href
  {https://ui.adsabs.harvard.edu/abs/2007MNRAS.375.1070B} {375, 1070}

\bibitem[\protect\citeauthoryear{Blandford \& Hernquist}{Blandford \&
  Hernquist}{1982}]{blandford1982magnetic}
Blandford R.,  Hernquist L.,  1982, Journal of Physics C: Solid State Physics,
  15, 6233

\bibitem[\protect\citeauthoryear{{Blandford} \& {Payne}}{{Blandford} \&
  {Payne}}{1982}]{Blandford82}
{Blandford} R.~D.,  {Payne} D.~G.,  1982, \mn@doi [\mnras]
  {10.1093/mnras/199.4.883}, \href
  {https://ui.adsabs.harvard.edu/abs/1982MNRAS.199..883B} {199, 883}

\bibitem[\protect\citeauthoryear{{Blandford} \& {Znajek}}{{Blandford} \&
  {Znajek}}{1977}]{Blandford77}
{Blandford} R.~D.,  {Znajek} R.~L.,  1977, \mn@doi [\mnras]
  {10.1093/mnras/179.3.433}, \href
  {https://ui.adsabs.harvard.edu/abs/1977MNRAS.179..433B} {179, 433}

\bibitem[\protect\citeauthoryear{Bodo, Ghisellini  \& Trussoni}{Bodo
  et~al.}{1992}]{bodo1992diamagnetic}
Bodo G.,  Ghisellini G.,   Trussoni E.,  1992, Monthly Notices of the Royal
  Astronomical Society, 255, 694

\bibitem[\protect\citeauthoryear{Broderick, Prakash  \& Lattimer}{Broderick
  et~al.}{2000}]{broderick2000equation}
Broderick A.,  Prakash M.,   Lattimer J.,  2000, The Astrophysical Journal,
  537, 351

\bibitem[\protect\citeauthoryear{{Broderick}, {Johannsen}, {Loeb}  \&
  {Psaltis}}{{Broderick} et~al.}{2014}]{Broderick14}
{Broderick} A.~E.,  {Johannsen} T.,  {Loeb} A.,   {Psaltis} D.,  2014, \mn@doi
  [\apj] {10.1088/0004-637X/784/1/7}, \href
  {https://ui.adsabs.harvard.edu/abs/2014ApJ...784....7B} {784, 7}

\bibitem[\protect\citeauthoryear{Caldarelli, Dias  \& Klemm}{Caldarelli
  et~al.}{2009}]{caldarelli2009dyonic}
Caldarelli M.~M.,  Dias O.~J.,   Klemm D.,  2009, Journal of High Energy
  Physics, 2009, 025

\bibitem[\protect\citeauthoryear{{Chatterjee}, {Elghozi}, {Novak}  \&
  {Oertel}}{{Chatterjee} et~al.}{2015}]{Chatterjee15}
{Chatterjee} D.,  {Elghozi} T.,  {Novak} J.,   {Oertel} M.,  2015, \mn@doi
  [\mnras] {10.1093/mnras/stu2706}, \href
  {http://adsabs.harvard.edu/abs/2015MNRAS.447.3785C} {447, 3785}

\bibitem[\protect\citeauthoryear{{Devlen} \& {Pek{\"u}nl{\"u}}}{{Devlen} \&
  {Pek{\"u}nl{\"u}}}{2007}]{Devlen07}
{Devlen} E.,  {Pek{\"u}nl{\"u}} E.~R.,  2007, \mn@doi [\mnras]
  {10.1111/j.1365-2966.2007.11677.x}, \href
  {https://ui.adsabs.harvard.edu/abs/2007MNRAS.377.1245D} {377, 1245}

\bibitem[\protect\citeauthoryear{{Devlen}, {Ulubay}  \&
  {Pek{\"u}nl{\"u}}}{{Devlen} et~al.}{2020}]{Devlen20}
{Devlen} E.,  {Ulubay} A.,   {Pek{\"u}nl{\"u}} E.~R.,  2020, \mn@doi [\mnras]
  {10.1093/mnras/stz3358}, \href
  {https://ui.adsabs.harvard.edu/abs/2020MNRAS.491.5481D} {491, 5481}

\bibitem[\protect\citeauthoryear{{Event Horizon Telescope Collaboration}
  et~al.,}{{Event Horizon Telescope Collaboration} et~al.}{2019}]{EHT19}
{Event Horizon Telescope Collaboration} et~al., 2019, \mn@doi [\apjl]
  {10.3847/2041-8213/ab0ec7}, \href
  {https://ui.adsabs.harvard.edu/abs/2019ApJ...875L...1E} {875, L1}

\bibitem[\protect\citeauthoryear{Felderhof}{Felderhof}{2004}]{felderhof2004relativistic}
Felderhof B.,  2004, The Journal of chemical physics, 120, 3598

\bibitem[\protect\citeauthoryear{{Fishbone} \& {Moncrief}}{{Fishbone} \&
  {Moncrief}}{1976}]{Fishbone76}
{Fishbone} L.~G.,  {Moncrief} V.,  1976, \mn@doi [\apj] {10.1086/154565}, \href
  {http://adsabs.harvard.edu/abs/1976ApJ...207..962F} {207, 962}

\bibitem[\protect\citeauthoryear{{Fragile}, {Gillespie}, {Monahan}, {Rodriguez}
   \& {Anninos}}{{Fragile} et~al.}{2012}]{Fragile12}
{Fragile} P.~C.,  {Gillespie} A.,  {Monahan} T.,  {Rodriguez} M.,   {Anninos}
  P.,  2012, \mn@doi [\apjs] {10.1088/0067-0049/201/2/9}, \href
  {https://ui.adsabs.harvard.edu/abs/2012ApJS..201....9F} {201, 9}

\bibitem[\protect\citeauthoryear{Ginzburg \& Syrovatsk}{Ginzburg \&
  Syrovatsk}{1969}]{ginzburg1969developments}
Ginzburg V.,  Syrovatsk S.,  1969, Annual Review of Astronomy and Astrophysics,
  7, 375

\bibitem[\protect\citeauthoryear{Goodman \& Xu}{Goodman \&
  Xu}{1994}]{goodman1994parasitic}
Goodman J.,  Xu G.,  1994, The Astrophysical Journal, 432, 213

\bibitem[\protect\citeauthoryear{Groot \& Suttorp}{Groot \&
  Suttorp}{1972}]{groot1972foundations}
Groot S. R.~d.,  Suttorp L.~G.,  1972

\bibitem[\protect\citeauthoryear{{Haardt} \& {Maraschi}}{{Haardt} \&
  {Maraschi}}{1991}]{Haardt91}
{Haardt} F.,  {Maraschi} L.,  1991, \mn@doi [\apjl] {10.1086/186171}, \href
  {https://ui.adsabs.harvard.edu/abs/1991ApJ...380L..51H} {380, L51}

\bibitem[\protect\citeauthoryear{{Hawley}, {Guan}  \& {Krolik}}{{Hawley}
  et~al.}{2011}]{Hawley11}
{Hawley} J.~F.,  {Guan} X.,   {Krolik} J.~H.,  2011, \mn@doi [\apj]
  {10.1088/0004-637X/738/1/84}, \href
  {https://ui.adsabs.harvard.edu/abs/2011ApJ...738...84H} {738, 84}

\bibitem[\protect\citeauthoryear{{Hawley}, {Richers}, {Guan}  \&
  {Krolik}}{{Hawley} et~al.}{2013}]{Hawley13}
{Hawley} J.~F.,  {Richers} S.~A.,  {Guan} X.,   {Krolik} J.~H.,  2013, \mn@doi
  [\apj] {10.1088/0004-637X/772/2/102}, \href
  {https://ui.adsabs.harvard.edu/abs/2013ApJ...772..102H} {772, 102}

\bibitem[\protect\citeauthoryear{{Hogg} \& {Reynolds}}{{Hogg} \&
  {Reynolds}}{2016}]{Hogg16}
{Hogg} J.~D.,  {Reynolds} C.~S.,  2016, \mn@doi [\apj]
  {10.3847/0004-637X/826/1/40}, \href
  {https://ui.adsabs.harvard.edu/abs/2016ApJ...826...40H} {826, 40}

\bibitem[\protect\citeauthoryear{{Huang}, {Huang}, {Rischke}  \&
  {Sedrakian}}{{Huang} et~al.}{2010}]{Huang10}
{Huang} X.-G.,  {Huang} M.,  {Rischke} D.~H.,   {Sedrakian} A.,  2010, \mn@doi
  [\prd] {10.1103/PhysRevD.81.045015}, \href
  {http://adsabs.harvard.edu/abs/2010PhRvD..81d5015H} {81, 045015}

\bibitem[\protect\citeauthoryear{{Jiang}, {Blaes}, {Stone}  \& {Davis}}{{Jiang}
  et~al.}{2019}]{Jiang19}
{Jiang} Y.-F.,  {Blaes} O.,  {Stone} J.,   {Davis} S.~W.,  2019, arXiv
  e-prints, \href {https://ui.adsabs.harvard.edu/abs/2019arXiv190401674J} {p.
  arXiv:1904.01674}

\bibitem[\protect\citeauthoryear{{Komissarov}}{{Komissarov}}{2006}]{Komissarov06}
{Komissarov} S.~S.,  2006, \mn@doi [\mnras] {10.1111/j.1365-2966.2006.10183.x},
  \href {https://ui.adsabs.harvard.edu/abs/2006MNRAS.368..993K} {368, 993}

\bibitem[\protect\citeauthoryear{Lesur \& Longaretti}{Lesur \&
  Longaretti}{2007}]{lesur2007impact}
Lesur G.,  Longaretti P.-Y.,  2007, Monthly Notices of the Royal Astronomical
  Society, 378, 1471

\bibitem[\protect\citeauthoryear{Maugin}{Maugin}{1978}]{Maugin:1978tu}
Maugin G.~A.,  1978, \mn@doi [J. Math. Phys.] {10.1063/1.523785}, 19, 1198

\bibitem[\protect\citeauthoryear{{McKinney}}{{McKinney}}{2006}]{McKinney06}
{McKinney} J.~C.,  2006, \mn@doi [\mnras] {10.1111/j.1365-2966.2006.10256.x},
  \href {https://ui.adsabs.harvard.edu/abs/2006MNRAS.368.1561M} {368, 1561}

\bibitem[\protect\citeauthoryear{M{\"o}sta, Ott, Radice, Roberts, Schnetter  \&
  Haas}{M{\"o}sta et~al.}{2015}]{mosta2015large}
M{\"o}sta P.,  Ott C.~D.,  Radice D.,  Roberts L.~F.,  Schnetter E.,   Haas R.,
   2015, Nature, 528, 376

\bibitem[\protect\citeauthoryear{{Novikov} \& {Thorne}}{{Novikov} \&
  {Thorne}}{1973}]{Novikov73}
{Novikov} I.~D.,  {Thorne} K.~S.,  1973, in Black Holes (Les Astres Occlus). pp
  343--450

\bibitem[\protect\citeauthoryear{Obergaulinger, Cerd{\'a}-Dur{\'a}n, M{\"u}ller
   \& Aloy}{Obergaulinger et~al.}{2009}]{obergaulinger2009semi}
Obergaulinger M.,  Cerd{\'a}-Dur{\'a}n P.,  M{\"u}ller E.,   Aloy M.~A.,  2009,
  Astronomy \& Astrophysics, 498, 241

\bibitem[\protect\citeauthoryear{{Oda}, {Machida}, {Nakamura}  \&
  {Matsumoto}}{{Oda} et~al.}{2009}]{Oda09}
{Oda} H.,  {Machida} M.,  {Nakamura} K.~E.,   {Matsumoto} R.,  2009, \mn@doi
  [\apj] {10.1088/0004-637X/697/1/16}, \href
  {https://ui.adsabs.harvard.edu/abs/2009ApJ...697...16O} {697, 16}

\bibitem[\protect\citeauthoryear{Pessah \& Goodman}{Pessah \&
  Goodman}{2009}]{pessah2009saturation}
Pessah M.~E.,  Goodman J.,  2009, The Astrophysical Journal Letters, 698, L72

\bibitem[\protect\citeauthoryear{{Pimentel} \& {Lora-Clavijo}}{{Pimentel} \&
  {Lora-Clavijo}}{2019}]{Pimentel2019}
{Pimentel} O.~M.,  {Lora-Clavijo} F.~D.,  2019, \mn@doi [\mnras]
  {10.1093/mnras/stz2750}, \href
  {https://ui.adsabs.harvard.edu/abs/2019MNRAS.490.4183P} {490, 4183}

\bibitem[\protect\citeauthoryear{{Pimentel}, {Lora-Clavijo}  \&
  {Gonzalez}}{{Pimentel} et~al.}{2018a}]{Pimentel20182}
{Pimentel} O.~M.,  {Lora-Clavijo} F.~D.,   {Gonzalez} G.~A.,  2018a, \mn@doi
  [\aap] {10.1051/0004-6361/201833736}, \href
  {https://ui.adsabs.harvard.edu/abs/2018A&A...619A..57P} {619, A57}

\bibitem[\protect\citeauthoryear{{Pimentel}, {Lora-Clavijo}  \&
  {Gonz{\'a}lez}}{{Pimentel} et~al.}{2018b}]{Pimentel20181}
{Pimentel} O.~M.,  {Lora-Clavijo} F.~D.,   {Gonz{\'a}lez} G.~A.,  2018b,
  \mn@doi [\apj] {10.3847/1538-4357/aac6d0}, \href
  {http://adsabs.harvard.edu/abs/2018ApJ...861..115P} {861, 115}

\bibitem[\protect\citeauthoryear{{Piran}}{{Piran}}{1999}]{Piran99}
{Piran} T.,  1999, \mn@doi [\physrep] {10.1016/S0370-1573(98)00127-6}, \href
  {https://ui.adsabs.harvard.edu/abs/1999PhR...314..575P} {314, 575}

\bibitem[\protect\citeauthoryear{{Porth} et~al.,}{{Porth}
  et~al.}{2019}]{Porth19}
{Porth} O.,  et~al., 2019, arXiv e-prints, \href
  {https://ui.adsabs.harvard.edu/abs/2019arXiv190404923P} {p. arXiv:1904.04923}

\bibitem[\protect\citeauthoryear{{Pringle}}{{Pringle}}{1981}]{Pringle81}
{Pringle} J.~E.,  1981, \mn@doi [\araa] {10.1146/annurev.aa.19.090181.001033},
  \href {https://ui.adsabs.harvard.edu/abs/1981ARA&A..19..137P} {19, 137}

\bibitem[\protect\citeauthoryear{Pu, Roy, Rezzolla  \& Rischke}{Pu
  et~al.}{2016}]{pu2016bjorken}
Pu S.,  Roy V.,  Rezzolla L.,   Rischke D.~H.,  2016, Physical Review D, 93,
  074022

\bibitem[\protect\citeauthoryear{Ray}{Ray}{2001}]{ray2001atomic}
Ray D.,  2001, Physical Review E, 63, 027401

\bibitem[\protect\citeauthoryear{Reboul-Salze, Guilet, Raynaud  \&
  Bugli}{Reboul-Salze et~al.}{2021}]{reboul2021global}
Reboul-Salze A.,  Guilet J.,  Raynaud R.,   Bugli M.,  2021, Astronomy \&
  Astrophysics, 645, A109

\bibitem[\protect\citeauthoryear{{Rees}}{{Rees}}{1984}]{Rees84}
{Rees} M.~J.,  1984, \mn@doi [\araa] {10.1146/annurev.aa.22.090184.002351},
  \href {https://ui.adsabs.harvard.edu/abs/1984ARA%26A..22..471R} {22, 471}

\bibitem[\protect\citeauthoryear{{Reeves}, {Pounds}, {Uttley}, {Kraemer},
  {Mushotzky}, {Yaqoob}, {George}  \& {Turner}}{{Reeves}
  et~al.}{2005}]{Reeves05}
{Reeves} J.~N.,  {Pounds} K.,  {Uttley} P.,  {Kraemer} S.,  {Mushotzky} R.,
  {Yaqoob} T.,  {George} I.~M.,   {Turner} T.~J.,  2005, \mn@doi [\apjl]
  {10.1086/498568}, \href
  {https://ui.adsabs.harvard.edu/abs/2005ApJ...633L..81R} {633, L81}

\bibitem[\protect\citeauthoryear{{Remillard} \& {McClintock}}{{Remillard} \&
  {McClintock}}{2006}]{Remillard06}
{Remillard} R.~A.,  {McClintock} J.~E.,  2006, \mn@doi [\araa]
  {10.1146/annurev.astro.44.051905.092532}, \href
  {https://ui.adsabs.harvard.edu/abs/2006ARA%26A..44...49R} {44, 49}

\bibitem[\protect\citeauthoryear{{Reynolds}}{{Reynolds}}{2014}]{Reynolds14}
{Reynolds} C.~S.,  2014, \mn@doi [\ssr] {10.1007/s11214-013-0006-6}, \href
  {https://ui.adsabs.harvard.edu/abs/2014SSRv..183..277R} {183, 277}

\bibitem[\protect\citeauthoryear{{Shakura} \& {Sunyaev}}{{Shakura} \&
  {Sunyaev}}{1973}]{Shakura73}
{Shakura} N.~I.,  {Sunyaev} R.~A.,  1973, \aap, \href
  {https://ui.adsabs.harvard.edu/abs/1973A&A....24..337S} {500, 33}

\bibitem[\protect\citeauthoryear{Singal}{Singal}{1986}]{singal1986magnetization}
Singal A.~K.,  1986, Astronomy and Astrophysics, 155, 242

\bibitem[\protect\citeauthoryear{{S{\k{a}}dowski}}{{S{\k{a}}dowski}}{2016}]{Sadowski16}
{S{\k{a}}dowski} A.,  2016, \mn@doi [\mnras] {10.1093/mnras/stw913}, \href
  {https://ui.adsabs.harvard.edu/abs/2016MNRAS.459.4397S} {459, 4397}

\bibitem[\protect\citeauthoryear{Suh \& Mathews}{Suh \&
  Mathews}{2010}]{suh2010magnetic}
Suh I.-S.,  Mathews G.~J.,  2010, The Astrophysical Journal, 717, 843

\bibitem[\protect\citeauthoryear{{Suzuki}, {Muto}  \& {Inutsuka}}{{Suzuki}
  et~al.}{2010}]{suzuki2010protoplanetary}
{Suzuki} T.~K.,  {Muto} T.,   {Inutsuka} S.-i.,  2010, \mn@doi [\apj]
  {10.1088/0004-637X/718/2/1289}, \href
  {https://ui.adsabs.harvard.edu/abs/2010ApJ...718.1289S} {718, 1289}

\bibitem[\protect\citeauthoryear{Wang, L{\"u}, Zhu  \& Wu}{Wang
  et~al.}{2016}]{wang2016diamagnetic}
Wang Z.,  L{\"u} G.,  Zhu C.,   Wu B.,  2016, Publications of the Astronomical
  Society of the Pacific, 128, 104201

\bibitem[\protect\citeauthoryear{{Wyithe} \& {Loeb}}{{Wyithe} \&
  {Loeb}}{2003}]{Wyithe03}
{Wyithe} J. S.~B.,  {Loeb} A.,  2003, \mn@doi [\apj] {10.1086/377475}, \href
  {https://ui.adsabs.harvard.edu/abs/2003ApJ...595..614W} {595, 614}

\makeatother
\end{thebibliography}

\bsp	\label{lastpage}
\end{document}